\documentclass[conference]{IEEEtran}

\usepackage{cite}
\usepackage{amsmath,amssymb,amsfonts}
\usepackage{algorithmic}
\usepackage{graphicx}
\usepackage{textcomp}
\usepackage{xcolor}
\usepackage{hyperref}
\usepackage{xurl}
\usepackage{booktabs}
\usepackage{listings}
\usepackage{subcaption}
\usepackage{tabularx}
\usepackage{pdflscape}
\usepackage{fancyhdr}

\usepackage[acronym,nomain,nonumberlist]{glossaries}

\lstdefinelanguage{yaml}{
    keywords={true,false,null,y,n},
    keywordstyle=\color{blue}\bfseries,
    basicstyle=\ttfamily,
    sensitive=false,
    comment=[l]{\#},
    morecomment=[s]{/*}{*/},
    morestring=[b]",
    morestring=[b]',
}

\newacronym{hpc}{HPC}{high performance computing}
\newacronym{qrmi}{QRMI}{quantum resource management interface}

\newcolumntype{L}{>{\raggedright\arraybackslash}X}

\hypersetup{
    colorlinks=true,
    linkcolor=black,
    filecolor=magenta,      
    urlcolor=cyan,
    citecolor=blue,
}

\newcommand{\ftt}[1]{{\footnotesize\ttfamily #1}}

\newcount\doubleblind
\newcount\newauthorlist
\title{Examining QRMI as a Unified Interface for Quantum-HPC Integration }

\author{
\ifnum\doubleblind>0
$<$omitted for double-blind review$>$
\else
\ifnum\newauthorlist>0
\IEEEauthorblockN{
Yoonho Park\IEEEauthorrefmark{4}
Munetaka Ohtani\IEEEauthorrefmark{4}
Emre M. Sahin\IEEEauthorrefmark{12}
Aleksander Wennersteen\IEEEauthorrefmark{8}\\
Daniel Gruber\IEEEauthorrefmark{3}
James Machin\IEEEauthorrefmark{1}
Mateusz Meller\IEEEauthorrefmark{12}
Doug Oucharek\IEEEauthorrefmark{7}
Vanessa Sochat\IEEEauthorrefmark{5}
Oscar Wallis\IEEEauthorrefmark{12}\\
Thomas Badts\IEEEauthorrefmark{8}
Tim Boyle\IEEEauthorrefmark{6}
Claudio Carvalho\IEEEauthorrefmark{4}
Antonio C\'orcoles\IEEEauthorrefmark{4}
Andrew Damin\IEEEauthorrefmark{10}
Vadim Elisseev\IEEEauthorrefmark{4}\\
Jonathan Frassineti\IEEEauthorrefmark{2}
Hiroshi Horii\IEEEauthorrefmark{4}
Eun-Kyung Lee\IEEEauthorrefmark{4}
Sara Marzella\IEEEauthorrefmark{2}
Daniel Milroy\IEEEauthorrefmark{5}
Matthieu Moreau\IEEEauthorrefmark{8} \\
Elisabeth Ortega-Carrasco\IEEEauthorrefmark{9}
Adarsh Patil\IEEEauthorrefmark{11}
G\'abor Samu\IEEEauthorrefmark{4}
Seetharami Seelam\IEEEauthorrefmark{4}
Amir Shehata\IEEEauthorrefmark{7}
James Thorne\IEEEauthorrefmark{6}
}
\else
\IEEEauthorblockN{
Thomas Badts\IEEEauthorrefmark{8}
Tim Boyle\IEEEauthorrefmark{6}
Claudio Carvalho\IEEEauthorrefmark{4}
Antonio C\'orcoles\IEEEauthorrefmark{4}
Andrew Damin\IEEEauthorrefmark{10}
Vadim Elisseev\IEEEauthorrefmark{4} \\
Jonathan Frassineti\IEEEauthorrefmark{2}
Daniel Gruber\IEEEauthorrefmark{3}
Hiroshi Horii\IEEEauthorrefmark{4}
Eun-Kyung Lee\IEEEauthorrefmark{4}
James Machin\IEEEauthorrefmark{1}
Sara Marzella\IEEEauthorrefmark{2} \\
Mateusz Meller\IEEEauthorrefmark{12}
Daniel Milroy\IEEEauthorrefmark{5}
Matthieu Moreau\IEEEauthorrefmark{8}
Munetaka Ohtani\IEEEauthorrefmark{4}
Elisabeth Ortega-Carrasco\IEEEauthorrefmark{9} \\
Doug Oucharek\IEEEauthorrefmark{7}
Yoonho Park\IEEEauthorrefmark{4}\IEEEauthorrefmark{13}
Adarsh Patil\IEEEauthorrefmark{11}
Emre M. Sahin\IEEEauthorrefmark{12}
G\'abor Samu\IEEEauthorrefmark{4}
Seetharami Seelam\IEEEauthorrefmark{4} \\
Amir Shehata\IEEEauthorrefmark{7}
Vanessa Sochat\IEEEauthorrefmark{5}
James Thorne\IEEEauthorrefmark{6}
Oscar Wallis\IEEEauthorrefmark{12}
Aleksander Wennersteen\IEEEauthorrefmark{8}
}
\fi

\IEEEauthorblockA{
\IEEEauthorrefmark{1}Alice \& Bob
\IEEEauthorrefmark{2}CINECA
\IEEEauthorrefmark{3}HPC Gridware
\IEEEauthorrefmark{4}IBM
\IEEEauthorrefmark{5}Lawrence Livermore National Laboratory \\
\IEEEauthorrefmark{6}National Quantum Computing Centre
\IEEEauthorrefmark{7}Oak Ridge National Laboratory
\IEEEauthorrefmark{8}Pasqal \\
\IEEEauthorrefmark{9}Qilimanjaro Quantum Tech
\IEEEauthorrefmark{10}Rensselaer Polytechnic Institute
\IEEEauthorrefmark{11}Siemens
\IEEEauthorrefmark{12}STFC Hartree Centre \\
}
\fi

}

\begin{document}
\maketitle

\thispagestyle{fancy}
\lhead{}
\rhead{}
\chead{}
\lfoot{\footnotesize{
SC26 Workshops, November 15-20, 2026, Chicago, Illinois, USA
\newline 979-8-3195-1221-5/26/\$31.00 \copyright 2026 IEEE}}
\rfoot{}
\cfoot{}
\renewcommand{\headrulewidth}{0pt}
\renewcommand{\footrulewidth}{0pt}

\begin{abstract}
The efficient and scalable integration of quantum resources into high-performance computing (HPC) environments requires standardized mechanisms for resource management, scheduling, and workflow orchestration across diverse and heterogeneous infrastructures. The Quantum Resource Management Interface (QRMI) addresses this challenge through a thin, vendor-agnostic middleware layer that provides standardized APIs for scheduling, executing, and monitoring quantum workloads while exposing quantum resources as first-class schedulable resources alongside CPUs and GPUs. Although previous work demonstrated QRMI integration with the Slurm workload manager, its applicability across other workload managers remained unexamined. This paper extends the validation of QRMI to a broad range of workload managers, including PBS, LSF, Grid Engine, Kubernetes, and the Flux Framework, encompassing traditional batch schedulers, a cloud-native orchestration platform, and a graph-based scheduler. We examine the integration patterns, implementation requirements, and scheduler-specific considerations associated with each environment and compare QRMI with alternative approaches to quantum resource integration. We demonstrate that QRMI provides a portable and flexible abstraction layer that minimizes scheduler-specific modifications while enabling consistent access to heterogeneous quantum resources across both on-premises and cloud environments.
\end{abstract}

\begin{IEEEkeywords}
Quantum Resources, HPC, Job Management, Resource Management, Workload Managers, QRMI, Integration
\end{IEEEkeywords}

\section{Introduction}

The convergence of quantum and HPC presents unique challenges in integrating quantum systems into existing HPC infrastructure. There is a critical need for standardized interfaces for quantum resource management, particularly as workflow orchestration becomes increasingly important in quantum-HPC integration. Current approaches suffer from fragmentation, making it difficult to deploy quantum computing capabilities across diverse HPC environments.

The Quantum Resource Management Interface (QRMI) introduces a thin and vendor-agnostic middleware layer that abstracts the control and management of quantum resources~\cite{qrmi}. It exposes standardized APIs for accessing, scheduling, executing, and monitoring quantum workloads, decoupling application logic and resource management from hardware-specific implementations. By treating quantum resources as first-class schedulable resources—alongside CPUs and GPUs—QRMI enables unified visibility and control across heterogeneous systems, while simplifying integration of quantum resources in HPC environments.

Within the broader context of quantum–HPC integration, QRMI plays a central role in enabling hybrid workflows envisioned for Quantum-Centric Supercomputing (QCSC)~\cite{ibm-qcsc}. As quantum and HPC systems evolve from loosely coupled offload models toward tightly integrated, co-designed platforms, the need for consistent abstractions across heterogeneous resources becomes critical. The QRMI approach has been validated with Slurm. The QRMI APIs allow on-premises and cloud-hosted quantum devices from a variety of hardware vendors to be incorporated into Slurm via extensible interfaces.

In this paper, we examine the QRMI approach across a wider variety of workload managers including PBS\textregistered{} (Portable Batch System)\footnote{The open source version, OpenPBS, was used in this study.}, IBM LSF\textregistered{} (Load Sharing Facility)\footnote{The free version, LSF Community Edition, was used in this study.}, Grid Engine\textregistered{}\footnote{Grid Engine originated from Sun Grid Engine and continued through the open-source Open Cluster Scheduler (used in this study)~\cite{ge-ocs} and the commercial Gridware Cluster Scheduler.}, Kubernetes (k8s), and the Flux Framework. This is, to the best of our knowledge, the first examination of a quantum resource interface spanning traditional batch schedulers (Slurm, PBS, LSF, Grid Engine), a container orchestrator for elastic cloud resources (k8s), and a hierarchical workload manager that uses a graph-based scheduler (Flux Framework). Building on the initial Slurm integration, this paper makes the following contributions:

\begin{itemize}
    \item Examination of QRMI integrations across five major workload managers: (1) Slurm, (2) PBS, (3) LSF, (4) Grid Engine, and (5) k8s;
    \item Analysis of integration patterns for batch schedulers and a container orchestrator;
    \item Quantum integration in a graph scheduler;
    \item Suggestions for improvements to QRMI and workload managers to facilitate quantum-HPC integration, practical deployment guidelines, and lessons learned.
\end{itemize}

Due to the dynamic nature of quantum resources, the most significant integration challenge is tracking quantum resource availability to inform workload manager schedulers. Addressing this integration challenge requires utilizing workload manager-specific mechanisms not originally intended for quantum resource management.

The remainder of this paper is organized as follows: Section~\ref{sec:background} provides background on HPC workload management systems, quantum resources integration challenges, and existing quantum-HPC integration approaches. Section~\ref{sec:architecture} describes the QRMI architecture and design. Section~\ref{sec:integrations} describes the QRMI integrations across different workload managers. Section~\ref{sec:analysis} presents an analysis of the workload manager integrations and lessons learned. Section~\ref{sec:future} discusses future work. Section~\ref{sec:conclusion} concludes.
\section{Background and Related Work}
\label{sec:background}

This section discusses the workload managers used in the examination, the quantum integration challenges, and other quantum integration approaches.

\subsection{HPC Workload Managers}

A workload manager (or batch system) provides the interface to manage resources and schedule workloads~\cite{cc-intro}. It includes the job manager, which receives new jobs and moves them through job lifecycle states, and the resource manager, which is responsible for tracking and monitoring hardware resources. The paper also refers to workflows which is a set of tasks with dependencies that need to be executed in a specific order for proper handling of exchanging inputs and outputs of data~\cite{cc-intro}. Workflow tools or managers must decide the granularity at which to map tasks to execution workload units, which are then handled by a workload manager.

The traditional workload managers, Slurm~\cite{slurm-jette2023}~\cite{slurm-doc}, PBS~\cite{pbs-nitzberg2004}~\cite{pbs-doc}, LSF~\cite{lsf-zhou1993}~\cite{lsf-doc}, and Grid Engine~\cite{ge-gentzsch2001}, were chosen for this study due to their popularity in HPC data centers. All provide extensibility through custom resource definitions and scheduler policies. All provide extensibility of lifecycle management through callback, plugin, or hook frameworks that modify a variety of job and system events.

While k8s is not primarily a workload manager, it is included in this evaluation due to its growing adoption in HPC environments for hybrid workflows and elastic resource provisioning across on-premises and cloud infrastructure~\cite{k8s-burns2016}~\cite{k8s-doc}. k8s provides declarative resource management and supports extensibility through custom resources (CRDs), operators, admission controllers, and scheduler plugins. The default k8s scheduler is used in this study.

Flux is a next-generation workload management framework designed for exascale and post-exascale computing~\cite{flux-ahn2014}~\cite{flux-doc}. It features a hierarchical architecture with flexible module and plugin systems and a graph-based scheduler. It is extensible through its plugin architecture, including Jobtap plugins that enable custom scheduling policies, job prioritization, dependency management, and resource management logic, as well as through site-defined resources. Flux also supports extensible lifecycle management through event-driven plugins and callbacks, allowing custom actions to be triggered at key job and system lifecycle events.

\subsection{Quantum Resources Integration Challenges}

Integrating quantum resources with existing workload managers introduces several architectural, operational, and user-experience challenges, stemming from the fundamental mismatch between classical HPC scheduling models and the behavior of quantum resources.

\subsubsection*{Abstraction mismatches} Existing workload managers expect local and relatively static resources, while quantum resources may be remote and more dynamic in nature with differing capabilities and access semantics. This requires additional configuration layers and administrator-managed mappings between logical resource names and physical endpoints, introducing complexity and potential inconsistency across environments. Until quantum computing matures, users must still reason about quantum resource selection and performance, which vary across systems and providers. Ensuring that applications remain quantum resource-agnostic while still exposing enough control for performance tuning is a delicate balance. \cite{qrmi} highlights that developing appropriate software abstractions is a key hurdle.

\subsubsection*{Scheduling limitations} Traditional schedulers are not designed for hybrid quantum–HPC workflows that require tight coordination across CPUs/GPUs and quantum resources. Quantum resources may have latency constraints, session semantics, or limited availability windows that can be incompatible with queue-based batch scheduling. As noted in~\cite{ibm-qcsc}, current systems often require manual orchestration of workloads, job coordination, and data movement, significantly reducing productivity as well as incurring unnecessary costs. This motivates the development of new integration layers such as QRMI. Ensuring correct co-scheduling, for example, aligning HPC compute with quantum execution windows, and avoiding duplicated queues or fragmented scheduling domains remain challenging.

\subsubsection*{Lifecycle integration} The initial integration of quantum resources into Slurm through QRMI and the SPANK (Slurm Plug-in Architecture for Node and job Kontrol) plugin model is described in~\cite{qrmi}. The integration required hooking into multiple stages of the job lifecycle (submission, initialization, execution, teardown) to acquire and release quantum resources and propagate tokens and environment variables. Of course, other workload managers have different plugin models, APIs, lifecycle semantics, and process models which determine how environment variables are propagated, creating additional integration complexity.

\subsubsection*{Deployment and operational complexity} While deployment and operational complexity will ease as quantum resources mature, the current complexity is a challenge. Workload manager integrations require plugins or hooks with different language requirements. System dependencies such as compilers, libraries, and workload manager headers lead to installation and maintainability issues. Administrators must manage credentials, endpoints, and environment variables for each quantum resource, which increases operational burden.

\subsection{Other Quantum-HPC Integration Approaches}

Driven by practical needs, several efforts to integrate quantum computers into HPC environments exist such as QMIO~\cite{qmio} and Pilot-Quantum~\cite{pilot-quantum}. There has also been prior work integrating quantum resources with cloud-native orchestration engines like k8s~\cite{qubernetes}~\cite{kubernetes-hybrid}. None of these efforts, in our view, provides a sufficiently comprehensive framework for quantum-HPC integration. In QMIO a quantum resource is represented by a quantum control node managed by Slurm. In the k8s integrations a quantum resource is represented by a k8s object. However, QMIO and the k8s integrations do not provide an interface that will allow quantum resources to be integrated into other workload managers. Pilot-Quantum does not address the challenges of integrating quantum computers with workload managers. Two larger systematic projects currently being developed are addressing the challenges of integrating quantum computers into HPC software stacks as we explain below.

QDMI (Quantum Device Management Interface)~\cite{qdmi}~\cite{qdmi-github} provides a device-level management interface with an implementation for IQM quantum computers and a Slurm integration~\cite{qdmi-iqm}~\cite{qdmi-iqm-repo}. The QDMI-Slurm integration is similar to the QRMI-Slurm integration described in~\cite{qrmi}. While QRMI focuses on integration with HPC environments through workload managers, QDMI focuses on compiler toolchain integration and monitoring. In many ways, QDMI and QRMI complement each other.

The openQSE (open Quantum-HPC Software Ecosystem) project~\cite{openqse} is an open-source, community-driven effort to define a reference architecture and common interface boundaries for the quantum-HPC software stack.

The goal of openQSE is not to replace existing SDKs, vendor stacks, or open-source efforts, but to define vendor-neutral layer boundaries to allow applications, runtimes, resource managers, and quantum resource implementations to interoperate. The effort is organized into working groups spanning architecture, workflows and use cases, compilation, and resource management.

To ground these interface definitions in working code, the project is developing QFw (Quantum Framework), an orchestration framework that integrates quantum resources into conventional HPC infrastructure\cite{ornl-qfw}. The resource management working group is using this testbed to build an understanding of what a quantum resource interface must provide. Both QRMI and QDMI have been integrated into QFw behind the same adapter layer, allowing identical workflows to be driven through each interface side by side against the same scheduler integration and the same quantum hardware. This comparison exposes where the two interfaces converge, where their semantics differ, and which capabilities users actually require. The findings feed back into the openQSE specification effort, creating a loop in which architectural decisions are validated in a working system before being proposed as community standards.


\section{QRMI Architecture and Design}
\label{sec:architecture}

\begin{figure}[t]
    \centering
    \includegraphics[width=2.9in]{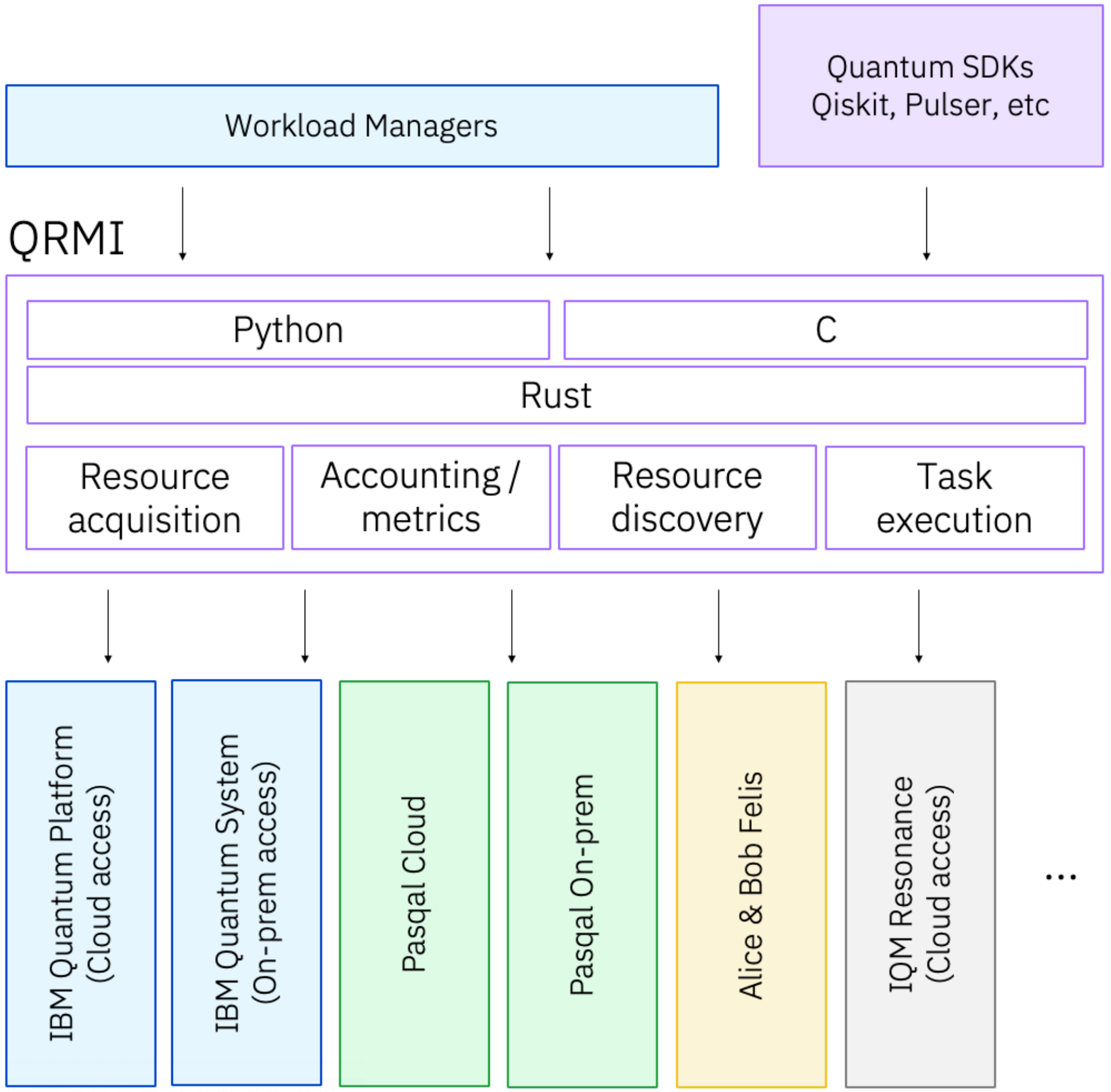}
    \caption{QRMI architecture, showing core language bindings, resource-management functions, and adapters connecting workload managers and quantum SDKs to heterogeneous quantum resources.}
    \label{fig:qrmi-arch}
    \vspace{-0.17in}
\end{figure}

As shown in Figure~\ref{fig:qrmi-arch}, QRMI serves as a vendor-agnostic abstraction layer that enables quantum computers to be managed as first-class schedulable resources within existing HPC workload management systems \cite{qrmi}. Users submit hybrid quantum–HPC jobs through a familiar workload manager interface. QRMI sits between the workload manager (such as Slurm) and diverse quantum resources. QRMI decouples workload manager logic from hardware-specific APIs, allowing the same operational workflow to support multiple quantum providers and deployment models. Through this abstraction, quantum resources can be represented consistently regardless of whether they correspond to a cloud-hosted quantum resource, a single QPU, a partition of a larger device, or a set of parallel execution lanes, simplifying administration and enabling unified scheduling, accounting, and access control.

QRMI addresses integration challenges by reshaping how quantum resources are exposed, scheduled, and consumed within existing systems. Its approach is a combination of standardization, indirection, and incremental integration.

\subsubsection*{Abstract heterogeneity} To address abstraction mismatches, QRMI abstracts heterogeneity by providing a uniform, vendor-neutral API for interacting with quantum resources. Instead of requiring workload managers and applications to understand provider-specific details, QRMI exposes a consistent interface for resource discovery, acquisition/release, job management, and observability. This decouples workload manager and application logic from hardware details and enables the same integration approach to work with multiple hardware providers. By shifting complexity into a well-defined middleware layer, QRMI allows workload managers and applications to integrate against the same abstraction rather than different quantum resource APIs.

\subsubsection*{Treat quantum resources as first-class schedulable resources} QRMI represents quantum devices as first-class resources within existing workload managers, similar to CPUs or GPUs. This helps address scheduling and co-scheduling challenges by integrating quantum resources into existing resource accounting and allocation models and enabling hybrid job submission through familiar interfaces. QRMI tries to reuse existing workload manager semantics, for example, Slurm GRES (Generic Resources), and extends them just enough to incorporate quantum execution. This is a pragmatic compromise that enables early and incremental adoption without requiring wholesale changes. This approach does have limitations which will be discussed later in the paper.

\subsubsection*{Use thin integration layers for lifecycle management} QRMI addresses lifecycle management through thin integration layers such as Slurm SPANK plugins that map lifecycle events to QRMI operations. At {\em job start} QRMI acquires quantum resources and tokens for each quantum resource. At {\em job execution} QRMI exposes connection information for each quantum resource via environment variables. At {\em job completion} QRMI releases quantum resources and associated tokens. This design preserves the native control flow, avoids invasive changes, and provides a clear contract between lifecycle management and quantum resource management. While this approach is not fully portable between workload managers due to differences in workload manager design, this approach does create a repeatable pattern (referred to as acquire-execute-release)  that will be discussed later in the paper.


\subsubsection*{Encapsulate complexity into configuration and middleware} QRMI pushes environment- and provider-specific complexity into configuration files (quantum resource definitions and credentials) and middleware logic (resource acquisition, authentication, session management). This addresses operational and heterogeneity challenges by keeping the scheduler interface simple and stable, isolating provider-specific logic from user workflows, and allowing administrators to centrally manage resource mappings. Although this increases configuration burden, it enables consistent behavior across environments and reduces coupling between application code and infrastructure.

\subsubsection*{Integration Challenges Not Discussed} As shown in Figure~\ref{fig:qrmi-arch}, QRMI provides resource discovery and accounting information. QRMI resource discovery allows users to specify constraints, for example, the number of qubits or the type of quantum processor. Some workload managers, such as LSF and Grid Engine, support the ability to define resource characteristics and specify resource constraints. QRMI accounting information makes quantum resource and quantum job metrics available in a QRMI database. Most workload managers provide accounting information. However, not all workload managers provide support for extending accounting information, for example, Slurm. Because both QRMI and workload managers provide resource discovery and accounting information, the challenge is defining the responsibilities of QRMI and the workload managers. This challenge is out of scope for this paper.

\section{QRMI Workload Manager Integrations}
\label{sec:integrations}

The integration of QRMI with Slurm introduced design patterns for incorporating quantum resources into established HPC environments \cite{slurm-qrmi}. Extending this work, we describe the integration of QRMI with Slurm~\cite{slurm-qrmi}, PBS~\cite{pbs-qrmi}, LSF~\cite{lsf-qrmi}~\cite{lsf-quantum}, and Grid Engine~\cite{ge-qrmi} with respect to four key design patterns: (1) define quantum resources as a custom resource, (2) use thin integration layers for lifecycle management, (3) encapsulate complexity into configuration and middleware, and (4) track quantum resource availability. We then describe the integration of QRMI with k8s and the Flux Framework. Although the k8s integration does not strictly follow these design patterns, it nevertheless supports quantum-HPC workloads. The Flux Framework integration remains incomplete but provided valuable insights into scheduling challenges. The section ends with a discussion of other integration considerations which are captured in Table \ref{tab:int-cons}.

\subsection{Define quantum resources as a custom resource}

All the workload managers support defining quantum resources as a custom resource. Below are explanations from the perspective of administrators and users.

\subsubsection*{Slurm} In~\cite{qrmi} quantum resources were defined as Generic Resources (GRES). However, GRES was not adopted because it represents node-bound resources and incorporating custom scheduling logic requires building plugin code within the Slurm source tree. Instead, we define the quantum resource as a license resource, leveraging Slurm’s Dynamic License mechanism ~\cite{slurm-dyn-licenses}, which allows an external process to control the license count at runtime independently of any specific node. An external daemon running on one cluster node acts as the license manager for the quantum resource. As shown in Listings~\ref{lst:slurm-add-license}, a license is defined via \ftt{sacctmgr}, and the daemon decrements its \ftt{lastconsumed} count from 1 to 0 once the quantum resource becomes ready to execute jobs. 

\begin{lstlisting}[caption={Defining a quantum resource license and managing availabilities with \ftt{sacctmgr}},label={lst:slurm-add-license}]
sacctmgr add resource name=ibm_fez count=1 \
    cluster=<your cluster name> (for example, linux) \
    allowed=100 type=license

# Quantum resource unavailable: consume the license
sacctmgr -i update resource ibm_fez set lastconsumed=1

# Quantum resource available: release the license
sacctmgr -i update resource ibm_fez set lastconsumed=0
\end{lstlisting}

The Slurm scheduler then allocates quantum and HPC resources together and transitions the job to the Execute state. Because the daemon relies on polling, this approach can suffer from a race condition in which the license count does not match the actual resource status. Then users can request quantum resources using simple command-line flags, for example, \ftt{sbatch --licenses=ibm\_fez@slurmdb:1}. This decouples the quantum application code from underlying cluster infrastructure configurations.

\subsubsection*{PBS} PBS requires two steps: (1) define the custom resource with \ftt{qmgr} and (2) make the scheduler aware of it~\cite{pbs-qrmi}. \ftt{qmgr} is the PBS administrator command-line interface. It sends management directives to the PBS server, such as creating queues, setting server or queue attributes, defining custom resources, and configuring nodes/vnodes. An example of defining a quantum resource with \ftt{qmgr} is shown in Listing~\ref{lst:pbs-add-res}. To make the PBS scheduler aware of the quantum resource, the resource must be added to the scheduler's \ftt{resources} list in \ftt{\$PBS\_HOME/sched\_priv/sched\_config} also shown in Listing~\ref{lst:pbs-add-res}. Users can then run jobs as usual with \ftt{qsub} with a specific quantum resource, for example, \ftt{qsub -l ibm\_fez=1 job.sh}.

\begin{lstlisting}[caption={Defining a quantum resource with \ftt{qmgr} and in the scheduler \ftt{resources} list.},label={lst:pbs-add-res}]
qmgr -c "create resource ibm_fez type=long, flag=nh"
qmgr -c "set node node001 resources_available.ibm_fez=1"

# resources in sched_config
resources: "ncpus, mem, arch, host, vnode, aoe, eoe, ibm_fez"
\end{lstlisting}

\subsubsection*{LSF} LSF provides a resource definition and resource-requirement model. Administrators can define resources in the \ftt{Resource} section of \ftt{\$LSF\_ENVDIR/lsf.shared} where each resource is assigned a name and type as shown in Listing \ref{lst:lsf-qpu}. The optional scheduling attribute (\ftt{qpu\_ready}) is explained below. Users can then run jobs as usual with \ftt{bsub} with a specific quantum resource through the \ftt{QRMI\_QPU\_RESOURCES} environment variable, for example, \ftt{QRMI\_QPU\_RESOURCES=ibm\_fez bsub job.sh}.

\begin{lstlisting}[caption={LSF quantum resource configuration.},label={lst:lsf-qpu}]
Begin Resource
RESOURCENAME  TYPE     INTERVAL  DESCRIPTION
ibm_fez       Boolean  ()        (IBM QPU name)
qpu_ready     Numeric  30        (QPU availability)
End Resource
\end{lstlisting}

\subsubsection*{Grid Engine} The Grid Engine family represents site-defined resources through complexes. In the QRMI integration~\cite{ge-qrmi}, quantum resources are modeled using a combination of resource-selection and resource-capacity complexes. The quantum resource identity is represented by a requestable, non-consumable string complex \ftt{qpu}. Users select a quantum resource using the standard resource request syntax, for example, \ftt{qsub -l qpu=PASQAL\_FRESNEL job.sh}. The scheduler dispatches the job only to execution hosts whose complex values advertise the requested quantum resource.
\subsection{Use thin integration layers for lifecycle management}

All the workload managers support customization of lifecycle management through callbacks, plugins, or hooks that can accommodate the acquire-execute-release pattern. The workload manager callbacks, plugins, and hooks used for the acquire-execute-release patterns are listed in Table~\ref{tab:int-cons}. This section describes the differences in the workload manager process structures which dictate where and when the callbacks, plugins, and hooks are run and how configuration information is passed between the acquire, execute, and release steps.

Slurm SPANK plugins provide lifecycle customization through callbacks. Because the SPANK callbacks used for acquire, execute, and release run in the same process on the execution node, the QRMI SPANK plugin stores configuration information as part of an internal context structure.

LSF and Grid Engine provide lifecycle customization through hooks. The acquire and execute steps run in one hook. The release step runs in another hook. Because the two hooks run in different processes on the execution node, configuration information is passed in a temporary file.

PBS provides lifecycle customization through hooks~\cite{pbs-hooks}. The acquire step runs in a hook executed in the PBS server daemon on a management node. The execute and release steps run in a MoM (Machine oriented Mini-server) daemon on the execution node. Configuration information is passed through the job’s variable list (\ftt{job.Variable\_List}).
\subsection{Encapsulating complexity into configuration and middleware}

To encapsulate complexity, users request quantum resources through standard submission interfaces using logical resource names and scheduler options. Administrators define the mapping from those logical names to physical quantum resources and credentials, and QRMI encapsulates the provider-specific logic needed to establish and manage access. Application code should remain as quantum resource-agnostic as possible. Application code should not handle session setup or quantum resource-specific configuration because those concerns are delegated to the workload manager configuration and QRMI.

The Slurm, PBS, LSF, and Grid Engine configuration is through the \ftt{qrmi\_config.json} file which is administrator-managed. \ftt{qrmi\_config.json} contains all the information that the integration layers needs to access the quantum resources, including API endpoints and credentials.


Although QRMI does not prescribe a specific credential-management model, secure handling of authentication data remains a critical deployment consideration. By default QRMI places credentials in \ftt{qrmi\_config.json} which is an administrator-owned file and credentials are passed through environment variables. While this is convenient, an adversarial, multi-tenant environment requires more secure management of credentials. For example, the \ftt{pasqal-local} QRMI implementation relies on MUNGE~\cite{munge} to generate authentication tokens scoped to the cluster. These tokens are decoded to extract a Linux UID which is then used for authorization checks. The Linux user base remains the source of truth for access control. As a result, the hosting site administrators do not need to maintain a separate access database for the on-premise quantum resources and no credentials are needed in \ftt{qrmi\_config.json}.
\subsection{Track quantum resource availability}

Classical batch schedulers assume that resources are relatively stable and continuously schedulable. Quantum resources, however, may become temporarily unavailable because of calibration and maintenance requirements, and access is often governed by provider-specific queueing and session models. As a result, workload managers require mechanisms that allow schedulers to track quantum resource availability and incorporate that information into scheduling decisions.

Existing approaches fall into two categories: (1) mechanisms that update resource availability independently of scheduling cycles and (2) mechanisms that evaluate resource availability during scheduling cycles. Slurm, LSF, and Grid Engine implement the former approach, whereas PBS implements the latter.

In Slurm, LSF, and Grid Engine, quantum resource availability is represented through scheduler-defined variables that are updated periodically by site-specific programs. These variables are then used by the scheduler when making placement decisions. Table~\ref{tab:track-availability} summarizes the mechanisms provided by each system. Because availability information is refreshed only periodically, the scheduler may make decisions using stale information if the state of a quantum resource changes between updates.

PBS addresses this limitation by evaluating resource availability at scheduling time. It provides two mechanisms that execute at the beginning of each scheduling cycle: (1) the \ftt{runjob} hook and (2) dynamic resource scripts. The \ftt{runjob} hook, which is also used during the acquire and execute phases, can check whether a quantum resource is currently available. If the resource is unavailable, the hook places the job in the pending state, and the check is repeated during the next scheduling cycle.

PBS also supports administrator-defined dynamic resources (\ftt{server\_dyn\_res}) configured through \ftt{\$PBS\_HOME/sched\_priv/sched\_config}. Each dynamic resource is associated with a script that executes during every scheduling cycle and reports the current availability of the corresponding resource.

The two PBS mechanisms are complementary. Because \ftt{runjob} is evaluated for every candidate job, it can result in redundant availability queries. Dynamic resource scripts, in contrast, perform a single availability query per quantum resource during each scheduling cycle. A combined approach therefore reduces duplicate checks while still allowing \ftt{runjob} to enforce resource availability constraints using the information collected by the dynamic resource scripts.

\begin{table*}[ht]
  \scriptsize
  \begin{tabularx}{\textwidth}{LLLL}
    \toprule
     & Slurm & LSF & Grid Engine \\
    \midrule
    Represent quantum resource availability & Use Dynamic Licenses~\cite{slurm-dyn-licenses} to represent quantum resource availability & Add attribute such as \ftt{qpu\_ready} to \ftt{Resource} section of \ftt{\$LSF\_ENVDIR/lsf.shared} & Define resource complexes with attributes such as \ftt{qpu\_ready} and \ftt{qpu\_slots} \\
    \midrule
    Track quantum resource availability & External daemon updates dynamic licenses representing quantum resources & ELIM program updates \ftt{qpu\_ready} for quantum resources & Load Sensor program updates \ftt{qpu\_ready} and \ftt{qpu\_slots} for quantum resources \\
    \bottomrule
  \end{tabularx}
  \caption{Tracking quantum resource availability with Slurm, LSF, and Grid Engine.}
  \label{tab:track-availability}
  \vspace{-0.17in}
\end{table*}

\subsection{k8s Operator Integration}

Because k8s is structured quite differently than traditional workload managers, the description of the k8s integration does not strictly follow the design patterns. However, the resulting k8s integration described here can support quantum--HPC workloads and demonstrates the flexibility of k8s and QRMI~\cite{k8s-qrmi}.

The k8s integration exposes quantum resources as schedulable resources via a combination of two custom k8s API resources. The first is the \ftt{QuantumResource} (QR), which is created by a cluster administrator and configures access to a particular quantum resource. The second is the \ftt{QuantumResourceClaim} (QRC), which represents a request for access to a particular QR. Valid QRCs create a k8s \ftt{Secret} to be consumed by a workload. The \ftt{Secret} contains data which provides access to the requested quantum resource.


There are two main advantages of the claim model. First, claims are resources in the k8s API server which gives them visibility to k8s API clients such as \ftt{kubectl}. The second advantage is in extensibility because a QRC itself is not tied to any particular workload type. These advantages provide users with more control over quantum-HPC workloads. However, a QRC could outlive the workload it was created for which could lead to unnecessary quantum resource acquisition time because the QRC is responsible for both the quantum resource acquire and release. A QRC that outlives the workload delays the release unnecessarily. The simple time-to-live parameter on QRCs is a preliminary mitigation. A more complete mitigation is the addition of \ftt{Jobs} annotations to support QRC. The operator creates a QRC from the annotations which allows the operator to associate the QRC with a particular job and enables the automatic cleanup of QRCs when the job is deleted or when the job finishes, whichever happens first. The association of QRC and \ftt{Job} prevents a possible race condition where the \ftt{Job} could start before the QRC is bound. The operator simply suspends all jobs with unbound claims.

The design adheres to the k8s-native approach of representing the problem-space ontology as k8s API resources through \ftt{CustomResourceDefinitions} (CRDs) which are implemented by an operator which itself runs as a workload in the cluster and captures all the logistical complexity. The \ftt{QuantumResource} CRD can be seen as equivalent to a resource block in \ftt{qrmi\_config.json} from the Slurm integration. The design aims to adhere to established k8s design patterns and conventions, in particular, in the implementation of a claim-style resource in the style of \ftt{storage.k8s.io}, in the use of resource annotations, and in the use of finalizers for cleanup. Note that this prototype contains no modification of the k8s scheduler itself. The only modification to scheduling logic leverages a feature particular to \ftt{Jobs}, allowing them to be suspended until claims are ready. This decision would likely be revisited in a more complete implementation should there be a desire to support tighter integration with a wider variety of workload types or to support more complex co-scheduling of resources.
\subsection{Flux Framework Integration}
\label{subsec:flux}

The Flux Framework integration has focused on the challenges of scheduling hybrid quantum-HPC workloads. This section describes the current state and the next steps needed for complete integration.

To treat quantum resources as a schedulable resource, quantum resources are modeled in the Fluxion resource graph.
While other workload managers consider quantum resources as job features, Flux models quantum devices as full-fledged hardware in the resource graph, on equivalent footing with HPC compute. When a user requests a job, they can request quantum and HPC resources. A job request can be represented as a resource ``shape'' that must be matched within the graph. Fluxion performs a depth-first search across both resources and time, returning a match when sufficient resources are available. Different match policies determine how matches are searched and scored, and how nodes are pruned from the graph. When a match is found for quantum and HPC resources, the resources are allocated for the duration of the job.


The modular design of the Flux Framework allows the scheduler to be integrated with other workload managers, or the Flux workload manager to use other schedulers. To avoid the complexity of modifying the Flux core modules, integration experiments were run using k8s as a workload manager. A k8s custom scheduler plugin, \emph{Fluence}, based on the Flux scheduler was developed from previous work~\cite{gromacs-kubeflux}~\cite{interfaces-kubeflux}~\cite{fluence-scalability} to model quantum and HPC resources in the resource graph~\cite{fluence}. Integration experiments enabled study of the two-queue problem which we describe briefly here and will be described in more detail in upcoming work \cite{sochat2026hybrid}.

The two-queue problem is the coordination of jobs in two queues, where one queue (quantum) does not provide transparency to running times. Given that HPC workload resources should execute at the same time as quantum, it becomes problematic if a quantum submission has an unknown waiting time. A conservative approach is to start HPC resources and incur the cost of keeping them idle while waiting for quantum resources. The Fluence approach is to monitor the quantum queue depth and dispatch classical resources when the quantum work is about to run.


The next integration step will move quantum resources to the Flux resource graph and the Flux Core~\cite{flux-core}. The Flux Core will  provide users with a rich set of tools. An early prototype integration~\cite{flux-qrmi} that served as a learning example focused on using Flux shell plugins. Other plugin types will be used for a more holistic integration.

\subsection{Other Integration Considerations}

The previous sections describe how workload managers address the key design patterns. Of course, the design patterns do not address all the integration considerations. Table~\ref{tab:int-cons} summarizes remaining considerations. Each row of Table~\ref{tab:int-cons} is described below. Note that the Flux Framework is not included in Table~\ref{tab:int-cons} due to its incomplete integration.

{\em Implementation language --} Extending workload manager functionality requires different language bindings for extension code. QRMI provides multiple language bindings, including Rust, C, and Python.
{\em Deployment --} Workload managers have different methods to distribute extension code across nodes. Some require extension code to be distributed and installed manually. Others provide support for distribution.
{\em Execution location --} Workload managers organize processes differently, which determines where extension code is executed.
{\em Lifecycle customization --} Which callbacks, plugins, and hooks provide QRMI operations acquire, execute, release.
{\em Job information access --} How extensions will access job information.
{\em Job rejection and modification --} How extensions will handle job rejection and modification.
{\em Environment variable passing --} How extensions can access, modify, and add environment variables to a job.
{\em Custom resource integration --} How new resources, such as quantum resources, can be added to the workload manager.
{\em Quantum resource availability --} Because quantum resources are typically shared between many users, submitting a job does not mean it will actually run in the near future. A workload manager must understand when a quantum resource is available or when a submitted job will actually run.
{\em Logging and debugging --} Where do extensions output information.
{\em Code update --} How are extensions updated.
{\em QHPC user experience --} How do users run quantum-HPC workloads.
\section{Analysis and Lessons Learned}
\label{sec:analysis}

This section discusses the support for design patterns across the workload managers examined in Section~\ref{sec:integrations}, credentials management, and lessons learned at QRMI deployments.

\subsection{Support for Design Patterns}
\newcommand{\fullsym}{\checkmark}
\newcommand{\nonesym}{$\times$}
\newcommand{\partialsym}{$\approx$}

In Section~\ref{sec:integrations}, we described how the workload managers support the QRMI design patterns that address the quantum integration challenges. Table~\ref{tab:design-patterns} summarizes the differences between the traditional workload managers: Slurm, PBS, LSF, and Grid Engine. k8s and the Flux Framework are not included.

\begin{table}[ht]
  \scriptsize
  \begin{tabularx}{\columnwidth}{Xcccc}
    \toprule
    Design pattern & Slurm & PBS & LSF & Grid Engine \\
    \midrule
    Define quantum resources as custom resources & \fullsym & \fullsym & \fullsym & \fullsym \\
    Use thin integration layers for lifecycle management & \fullsym & \fullsym & \fullsym & \fullsym \\
    Encapsulate complexity into configuration and middleware & \fullsym & \fullsym & \fullsym & \fullsym \\
    Track quantum availability & \partialsym$^{**}$ & \partialsym & \partialsym & \partialsym \\
    Native quantum resource support & \nonesym & \nonesym & \nonesym & \nonesym \\
    Overall & \partialsym & \partialsym & \partialsym & \partialsym \\
    \bottomrule
  \end{tabularx}
  \caption{Workload manager support for design patterns. \fullsym~is full support, \partialsym~is partial support, and \nonesym~is no support.}
  \label{tab:design-patterns}
  \vspace{-0.17in}
\end{table}

Full support of the first design pattern (treat quantum resources as schedulable resources) requires the ability to define new custom resources such as quantum resources and the ability to track the dynamic availability of quantum resources. While all the workload managers allow new custom resources to be defined, they all have limitations tracking the dynamic availability of quantum resources. Due to these limitations, they are all assigned partial support.

Slurm (specifically, GRES) cannot reflect the dynamic availability of quantum resources, hence, the special symbol (\partialsym$^{**}$) in Table~\ref{tab:design-patterns}. Slurm introduced Hierarchical Resources (HRES) in version 25.05 (May 2025)~\cite{slurm-hres}. HRES allows for license-like resources to be defined and associated with specific nodes. Jobs may request any integer count of that resource. If license counts could be dynamically adjusted to reflect quantum resource availability, Slurm HRES could represent quantum resources without the need for an external daemon. At present, however, this functionality is not supported.

While PBS, LSF, and Grid Engine can track quantum resource availability, there are still concerns about race conditions and scalability that must be carefully explored. Although Flux is not included in Table~\ref{tab:design-patterns} because its QRMI integration remains incomplete, the Fluxion approach described in Section~\ref{subsec:flux} offers an alternative mechanism for representing dynamic resource availability. The Fluxion approach is to monitor the length of the queue that contains quantum jobs. While the queue length of quantum jobs can be misleading, this is a practical solution absent additional information about the dynamic availability of quantum resources.

It is worth noting that the PBS approach to dynamic availability is different than the polling approach of LSF and Grid Engine. The PBS approach allows quantum resource availability to be checked in two places: (1) administrator-supplied script at each scheduling cycle for dynamic resources and (2) the \ftt{runjob} hook. The \ftt{runjob} hook is a viable alternative because it is executed at the PBS server before any other resources are allocated. An inaccessible quantum resource causes the hook to reject the job, leaving it pending in the queue rather than leaving other resources idle. These approaches are complementary. The dynamic resources script can filter out jobs before invoking the \ftt{runjob} hook which could be important as the number of concurrently pending jobs for the same quantum resource grows.

While it is important for workload managers to reflect the dynamic availability of quantum resources, it is also important that QRMI supports a vendor-agnostic method to reflect availability. QRMI has \ftt{is\_accessible()} as a quantum resource method, but this method does not reflect availability. The QRMI community is considering adding a \ftt{is\_runnable()} method to provide a time estimate for when a quantum job can expect to run.


``Native quantum resource support'' is a product support and maintenance statement. Some HPC data centers cannot incorporate non-product software in their production systems. None of the traditional workload managers have ``native quantum resource support'' which would exclude them from these HPC data centers. Of course, QRMI itself would be excluded as well. Addressing this requires collaboration between the QRMI community and workload manager producers.


\subsection{Deployments}
QRMI has been deployed at multiple sites, including
\ifnum\doubleblind>0
$<$omitted for double-blind review$>$
\else
CINECA~\cite{cineca-pasqal}, BasQ (Basque Quantum)~\cite{basq-ibm}, RPI (Rensselaer Polytechnic Institute)~\cite{rpi-ibm}, and the United Kingdom's HPC ecosystem~\cite{mary_coombs}~\cite{nqcc_testbeds}. CINECA is using QRMI with the Slurm SPANK plugin integration to integrate the Pasqal quantum resource ``Sol'' with its  production, EuroHPC Tier-0 system Leonardo. BasQ and RPI use QRMI with the Slurm SPANK plugin integration to integrate IBM quantum systems with existing HPC infrastructure. BasQ integrated an IBM Quantum System Two with a network of Basque research centers. RPI integrated an IBM Quantum System One with its AIMOS supercomputer.
\fi

QRMI's ability to access both cloud and on-premise quantum resources allowed 
\ifnum\doubleblind>0
$<$omitted for double-blind review$>$
\else
CINECA 
\fi
to begin integration work even before the quantum resource was available. The 
\ifnum\doubleblind>0
$<$omitted for double-blind review$>$
\else
CINECA 
\fi
deployment also illustrates the need for middleware beyond the workload manager, for example, by optimizing quantum resource usage with a second-level scheduler~\cite{elephant}, that is integrated with the system scheduler~\cite{pasqal-on-prem}. This further reinforces the value of QRMI's lightweight, middleware-centric design. Integrating QRMI with a large production HPC such as
\ifnum\doubleblind>0
$<$omitted for double-blind review$>$
\else
Leonardo,
\fi
also surfaced other features to improve within QRMI, such as ensuring that the logs were correctly routed to the admin-defined Slurm logs, security, and setup of the integration in a situation where there are many stakeholders, such as the system administrators, HPC-QC, and the vendor.

The
\ifnum\doubleblind>0
$<$omitted for double-blind review$>$
\else
RPI
\fi
integration uncovered the quantum-HPC co-scheduling problem. 
\ifnum\doubleblind>0
$<$omitted for double-blind review$>$
\else
AIMOS
\fi
supercomputer GPUs were left idle for long periods of time due to hybrid quantum-HPC workloads getting allocated to GPUs while waiting for a quantum resource. One possible solution is to split quantum-HPC workloads into separate quantum and HPC jobs as suggested in~\cite{slurm-hetero}. However, this requires users to change how they behave. As discussed previously, there are multiple ways to address the quantum-HPC co-scheduling problem without requiring users to change their behavior.

Within the 
\ifnum\doubleblind>0
$<$omitted for double-blind review$>$
\else
United Kingdom's
\fi
HPC ecosystem, QRMI has been deployed on development test nodes of the
\ifnum\doubleblind>0
$<$omitted for double-blind review$>$
\else
Mary Coombs
\fi
supercomputer at the
\ifnum\doubleblind>0
$<$omitted for double-blind review$>$
\else
STFC Hartree Centre,
\fi
where it is currently evaluated using cloud-accessible quantum resources. This environment supports the implementation and validation of new features, assessment of scalability, development of tests for early software versions, and investigation of prospective capabilities, including improved accounting support and expanded unit-test coverage. In parallel, the
\ifnum\doubleblind>0
$<$omitted for double-blind review$>$
\else
UK National Quantum Computing Centre (NQCC)
\fi
is undertaking a preliminary integration of its API into QRMI to enable access to compatible quantum systems, including on-premises testbeds and cloud-accessible resources that expose the supported interfaces of QRMI. This API integration establishes the technical foundation for subsequent deployment across
\ifnum\doubleblind>0
$<$omitted for double-blind review$>$
\else
NQCC
\fi
infrastructure, and extended evaluation of different quantum systems within a QCSC context.
\section{Future Work}
\label{sec:future}

Future work includes evolving the QRMI interface, refining current integrations, evolving QRMI resource discovery and accounting information, integration with workflow managers, and quantitative analysis of integrations.

\subsubsection*{QRMI interface} The QRMI community is working with the openQSE and QDMI communities to understand HPC data center requirements. While QRMI has been successful in integrating quantum resources in HPC data centers, quantum technologies are rapidly evolving and the QRMI interface must evolve with them.

\subsubsection*{Refining current integrations} Many of the integrations presented in this paper remain at the proof-of-concept stage. The QRMI community had limited contact with the workload manager developers. The QRMI community intends to collaborate more closely with workload manager developers to refine the current integrations and to eventually see native support for quantum resources in workload managers.

\subsubsection*{Resource discovery and accounting information} While it is possible to implement resource discovery and accounting at the workload manager, QRMI offers the possibility of providing these services consistently across workload managers. Resource discovery is challenging because quantum technologies expose different capabilities, while schedulers use different resource models and selection mechanisms. Future work will define abstractions that can be mapped to scheduler-specific models. Reliable accounting integration is also essential because HPC facilities use different accounting systems. QRMI therefore requires abstractions that represent quantum resource usage consistently across cloud and on-premises resources.

\subsubsection*{Workflow managers} Integrating QRMI with workflow managers such as Airflow~\cite{airflow}, Prefect~\cite{prefect}, Nextflow~\cite{nextflow}, and Snakemake~\cite{snakemake} is a key step toward scalable orchestration of hybrid quantum-HPC workloads. As quantum resources become part of larger scientific workflows, workflow managers require a consistent and reliable interface such as QRMI for interacting with diverse quantum resources. The workflow described in~\cite{qc-merz25} illustrates the complexity and scale that can benefit from QRMI integration with workflow managers. The workflow decomposes very large protein-ligand systems into embedded quantum fragments, samples fragment electronic configurations on quantum processors, reconstructs fragment wavefunctions via optimized subspace diagonalization on classical supercomputers, and then combines the fragment results to achieve coupled-cluster-level quantum chemistry accuracy for systems exceeding 12,000 atoms.

%

\subsubsection*{Quantitative analysis} The analysis presented in this paper is qualitative. While this is valuable because it helps the larger community understand the challenges of integrating quantum resources into HPC data centers, a quantitative analysis would tell the larger community more.
\section{Conclusions}
\label{sec:conclusion}

This paper examined QRMI across a diverse set of workload managers. We identified four design patterns: (1) define quantum resources as a custom resource, (2) use thin integration layers for lifecycle management, (3) encapsulate complexity into configuration and middleware, and (4) track quantum resource availability. The examination demonstrates that QRMI provides a portable and vendor-agnostic abstraction that can be integrated into substantially different scheduling and orchestration environments without requiring invasive modifications.

Our examination also highlights limitations of current workload managers when applied to quantum resources. While all evaluated systems provide mechanisms for extending scheduling and lifecycle management, support for dynamic quantum resource availability varies. Challenges such as provider-managed queues, calibration downtime, opaque execution latencies, and quantum-HPC co-scheduling can reduce overall system utilization if not addressed.
These findings suggest that future workload managers and orchestration systems will require stronger support for dynamic and heterogeneous resources than is typically needed in current HPC environments.

Overall, the examination supports the use of QRMI as a practical foundation for integrating heterogeneous quantum resources into emerging QCSC platforms.
Continued collaboration among hardware vendors, middleware developers, scheduler maintainers, and HPC data centers will be essential for evolving these proof-of-concept integrations into production-ready quantum-HPC ecosystems.

\section*{Acknowledgements}

\ifnum\doubleblind>0
$<$omitted for double-blind review$>$
\else
Authors were ordered according to contributions across five areas: writing, QRMI, workload manager integrations, data center integrations, and strategy. Authors contributing to four or five areas were placed in the first line, led by Yoonho Park. Those contributing to two or three areas in the second line. Those contributing to one area in the remaining lines. Within the contribution tiers, authors are listed alphabetically.

The authors thank Leonardo Torretta (Siemens), Stefano Mensa (NVIDIA), Zohim Chandani (NVIDIA), and Edoardo Altamura (NQCC) for fruitful discussions and editing.

Prepared by LLNL under Contract DE-AC52-07NA27344 (LLNL-CONF-2023303). This manuscript has been authored by Lawrence Livermore National Security, LLC under Contract No. DE-AC52-07NA27344 with the US. Department of Energy. The United States Government retains, and the publisher, by accepting the article for publication, acknowledges that the United States Government retains a non-exclusive, paid-up, irrevocable, world-wide license to publish or reproduce the published form of this manuscript, or allow others to do so, for United States Government purposes.

This work was supported by the Hartree National Centre for Digital Innovation, a collaboration between the Science and Technology Facilities Council and IBM.
\fi

The authors used AI-assisted tools, specifically, ChatGPT 5.5, Codex 5.6, and IBM Bob, during the preparation of this work for support tasks. All AI-generated content was reviewed, edited, and verified by the authors. The authors take full responsibility for the accuracy, integrity, and originality of the submitted work. AI tools were not used for analysis or the drawing of conclusions.

\begin{landscape}
\begin{table}
  \centering
  \scriptsize
  \begin{tabularx}{\linewidth}{>{\raggedright\arraybackslash}p{2cm}LLLLL}
    \toprule
     & {\bf Slurm} & {\bf PBS} & {\bf LSF} & {\bf Grid Engine} & {\bf k8s Operator}
    \\
    \midrule
    Implementation language & C (shared library \ftt{.so}) & Python (script) & Programming language agnostic & Programming language agnostic. Golang is preferred. & Rust
    \\
    \midrule
    Deployment & Compiled \ftt{.so} distributed to each node manually & Registered via {\tt qmgr}; server distributes automatically & \ftt{jobstarter} and \ftt{postexec} LSF extensions/plug-ins registered in \ftt{lsb.queues}; placed in \ftt{LSF\_SERVERDIR} & Deploy queue hooks binaries then configure complex entries via \ftt{qconf} & Operator that contains 2 custom resources: \ftt{QuantumResource} (for execute) and \ftt{QuantumResourceClaim} (for acquire-release)
    \\
    \midrule
    Execution location & \ftt{slurmctld} (controller) or \ftt{slurmd} (compute node), depending on context & PBS Server or MOM (compute node), selected per hook type & \ftt{jobstarter} and \ftt{postexec} run on (first) execution host & Queue \ftt{prolog} and \ftt{epilog} execute on the execution host; scheduling logic on master node & Claim processed on service node; execution occurs on worker node
    \\
    \midrule
    Job information access & SPANK API, for example, {\tt spank\_get\_item} to read C structs & {\tt pbs.event().job} object — attributes readable and writable directly & Parameter file and environment variables for submission hooks and execution hooks & Environment variables and command-line arguments provided by the execution daemon (\ftt{sge\_execd}) & NA
    \\
    \midrule
    Job rejection and modification & [Limited] Return an error code to abort; modifying attributes is difficult & [Flexible] {\tt event.reject()} and attribute rewriting are straightforward & Job error codes are returned in case of a {\tt jobstarter} or {\tt postexec} failure & Job Submission Verifier (JSV) framework can reject and modify job parameters and environment & NA
    \\
    \midrule
    Lifecycle customization & Acquire: \ftt{init\_post\_opt()} Execute: \ftt{task\_init()} Release: \ftt{exit()} & Acquire+execute: \ftt{runjob} Release: \ftt{execjob\_end} & Acquire+execute: \ftt{jobstarter} Release: \ftt{postexec} & Acquire+execute: \ftt{prolog} Release: \ftt{epilog} & NA
    \\
    \midrule
    Environment variable passing & {\tt spank\_\{set,get\}env} for direct manipulation & Via {\tt job.Variable\_List}; accessible as {\tt PBS\_VAR} inside the job script & Environment  variables accessible in hooks & Environment variables accessible in hooks & ``Inline'' or sourced from k8s Secrets
    \\
    \midrule
    Custom resource integration & Add SPANK options to {\tt sbatch} and combine them with licenses & Define custom resources via {\tt qmgr}; specify in select or {\tt -l}; read inside the hook & Add resource in the \ftt{Resource} section of \ftt{lsf.shared} & Resource ID and consumption separated; defined as STRING complexes & Defined through custom operator
    \\
    \midrule
    Quantum resource availability & Use licenses to represent availability; add external daemon to update license count & Can use \ftt{runjob} hook or script invoked by dynamic resources (\ftt{server\_dyn\_res}) & Extend LIM (Load Information Manager) via {\tt ELIM} hooks to update availability & Use Load Sensors scripts to update availability & Use sidecar containers to update availability$\ddagger$
    \\
    \midrule
    Logging and debugging & \ftt{slurm\_info}, \ftt{slurm\_error} $\rightarrow$ slurmd log & \ftt{pbs.logmsg()} $\rightarrow$ PBS server log (configurable log level) & \ftt{jobstarter} can write to \ftt{stdout,stderr,file}; \ftt{postexec} limited to \ftt{file} & Hook status exposed through job environment variables & NA
    \\
    \midrule
    Code update & C source must be rebuilt and redeployed on every change & Re-register the updated Python script with {\tt qmgr} & Redeploy scripts/binaries in {\tt LSF\_SERVERDIR} & Rebuild and redeploy hooks & Load updated container
    \\
    \midrule
    QHPC user experience & \ftt{sbatch --qpu=ibm\_fez --licenses=ibm\_fez@ job.sh} & \ftt{qsub -l quantum\_res=ibm\_fez job.sh} & \ftt{QRMI\_QPU\_RES=ibm\_fez bsub job.sh} & \ftt{qsub -l qpu=ibm\_fez job.sh} & \ftt{kubectl apply -f job.yaml}
    \\
    \bottomrule
  \end{tabularx}
  \caption{Summary of integration considerations. Note 1: To fit in the table, the Slurm SPANK callbacks were shortened by dropping the \ftt{slurm\_spank\_} prefix, the Slurm \ftt{--licenses=ibm\_fez@slurmdb:1} was shortened to \ftt{--licenses=ibm\_fez@}, the environment variable \ftt{QRMI\_QPU\_RESOURCES} was shortened to \ftt{QRMI\_QPU\_RES}, and the PBS \ftt{quantum\_resources} command line argument was shortened to \ftt{quantum\_res}. Note 2: $\ddagger$ indicates not implemented.}
  \label{tab:int-cons}
\end{table}
\end{landscape}
\bibliographystyle{IEEEtran}

\end{document}